\documentclass{llncs}
\usepackage{amssymb}
\usepackage{amsmath}
\usepackage{times}
\usepackage{comment}
\usepackage{graphicx}

\title{No small nondeterministic read-once branching programs for CNFs of bounded treewidth}
\author{Igor Razgon}
\institute{Department of Computer Science and Information Systems, Birkbeck, University of London\\
           igor@dcs.bbk.ac.uk}
\begin{document}
\maketitle
\begin{abstract}
In this paper, given a parameter $k$, we demonstrate an infinite class of {\sc cnf}s of treewidth at
most $k$ of their primary graphs such that the equivalent nondeterministic read-once branching programs 
({\sc nrobp}s) are of size at least $n^{ck}$ for some universal constant $c$. 
Thus we rule out the possibility of fixed-parameter space complexity of {\sc nrobp}s 
parameterized by the smallest treewidth of the equivalent  {\sc cnf}.   
\end{abstract}
\section{Introduction}
Read-once Branching Programs ({\sc robp}s) is a well known representation of Boolean functions. 
Oblivious {\sc robp}s, better known as  Ordered Binary Decision Diagrams ({\sc obdd}s),
is a subclass of {\sc robp}s, very well known because of its applications in the area of verification \cite{OBDD}.
An important procedure in these applications is transformation of a {\sc cnf} into an equivalent {\sc obdd}.
The resulting {\sc obdd} can be exponentially larger than the initial {\sc cnf}, however a space efficient
transformation is possible for special classes of functions. For example, it has been shown in \cite{VardiTWD}
that a {\sc cnf} of treewidth $k$ of its primal graph can be transformed into an {\sc obdd} of size $O(n^k)$.
A natural question is if the upper bound can be made fixed-parameter i.e. of the form $f(k)n^c$ for some constant $c$.
In \cite{RazgonKR} we showed that it is impossible by demonstrating that for each sufficiently large $k$
there is an infinite class of {\sc cnf}s of treewidth at most $k$ whose smallest {\sc obdd} is of size at least 
$n^{k/5}$.

In this paper we report a follow up result showing that essentially the same lower bound holds for
Non-deterministic {\sc robp}s ({\sc nrobp}s). In particular we show that there is a constant $0<c<1$
such that for each sufficiently large $k$ 
there is an infinite class of {\sc cnf}s of treewidth at most $k$ (of their primary graphs)
for which the space complexity of the equivalent {\sc nrobp}s  is at least $n^{ck}$. 
Note that {\sc nrobp}s are strictly more powerful than {\sc robp}s
in the sense that there is an infinite class of functions having a poly-size {\sc nrobp} representation 
and exponential {\sc robp} space complexity \cite{Yukna}. In the same sense, {\sc robp}s are strictly more powerful 
than {\sc obdd}s, hence the result proposed in this paper is a significant enhancement of
the result of \cite{RazgonKR}. 

We believe this result is interesting from the parameterized complexity theory perspective because
it contributes to the understanding of parameterized \emph{space} complexity of various representations
of Boolean functions. In particular, the proposed result implies that {\sc robp}s are inherently incapable to efficiently
represent functions that are representable by {\sc cnf}s of bounded treewidth. A natural question for further research
is the space complexity of read $c$-times branching programs \cite{Razktimes} (for an arbitrary constant $c$ 
independent on $k$) w.r.t. the same class of functions. 

\begin{comment}
We believe this result is also interesting from the parameterized complexity theory perspective. Indeed, treewidth is a major 
parameter for design of fixed-parameter algorithms, both in general and for the {\sc sat} problem in particular. 
Thus it is interesting to see what computational power is needed to properly utilize the small treewidth.
The result of this paper shows that {\sc robp}s (even without the determinism assumption) are inherently weak for this
purpose.
\end{comment}

To prove the proposed result, we use monotone $2$-{\sc cnf}s (their clauses are of form $(x_1 \vee x_2)$ where
$x_1$ and $x_2$ are $2$ distinct variables). These {\sc cnf}s are in one-to-one correspondence with graphs
having no isolated vertices: variables correspond to vertices and $2$ variables occur in the same clause if and only if the
corresponding vertices are adjacent. This correspondence allows us to use these {\sc cnf}s and graphs interchangeably.
\begin{comment}
To measure the complexity of graphs, we introduce the notion of Distant Matching Width ({\sc dmw}) of a graph $G$ denoted
by $dmw(G)$. In particular, if $dmw(G) \geq t$ then any permutation of $V(G)$ has a partition into a prefix $V_1$ and a suffix
$V_2$ such that there is a matching $M$ of size $t$ consisting of edges with one end in $V_1$ and the other end in $V_2$ with the ends of
different edges of $M$ being not adjacent and not having common neighbours. 
\end{comment}
We introduce the notion of Distant Matching Width ({\sc dmw}) of a graph $G$ and prove $2$ theorems. One of them states that
a {\sc nrobp} equivalent to a monotone 2-{\sc cnf} with the corresponding graph $G$ having {\sc dmw} at least $t$ is of size at least $2^{t/a}$ where $a$
is a constant dependent on the max-degree of $G$. The second theorem states that for each sufficiently large $k$ there is an
infinite family of graphs of treewidth $k$ and max-degree $5$ whose {\sc dmw} is at least $b*log n*k$ for some constant $b$ independent of $k$. 
The main theorem immediately follows from replacement of $t$ in the former lower bound by the latter one.

The strategy outlined above is similar to that we used in \cite{RazgonKR}. However, there are two essential differences.
First, due to a much more `elusive' nature of {\sc norbp}s compared to that of {\sc obdd}, the counting argument is more
sophisticated and more restrictive: it applies only to {\sc cnf}s whose graphs are of constant degree. Due to this latter
aspect, the target set of {\sc cnf} instances requires a more delicate construction and reasoning.

%This weakness constrasts with the Turing Machine model, where treewidth is a major parameter for 
%design of fixed-parameter algorithms for analysis of {\sc cnf}s, see \cite{SzPau} for a recent result in this direction 
%and for a nice overview of the previous results. 

\begin{comment}
A lot of research has been done on {\sc robp}s \cite{Yukna,Wegbook}. A modified version of the {\sc robp} has been useful for proof complexity 
\cite{RegRes,RaWiYao}. A recent result \cite{IwamaTE} shows a space $\Omega(p^h)$ lower bound 
of $p$-way deterministic {\sc robp}s for the Tree Evaluation Problem ({\sc tep}) \cite{TreeEval},
where $h$ is the height of the underlying tree and $\{1, \dots, p\}$ are the values that can be received
by its leaves. (The publications on {\sc tep} use $k$ to denote the number 
of values; we used $p$ to avoid confusion with $k$ as in the rest of this section). 
\end{comment}

Due to the space constraints, some proofs are either omitted or replaced by sketches. 
The complete proofs are provided in the appendix. 
%Structure of the paper
\section{Preliminaries}
In this paper by a \emph{set of literals} we mean one that does not
contain an occurrence of a variable and its negation.
For a set $S$ of literals we denote by $Var(S)$ the set of variables
whose literals occur in $S$. If $F$ is a Boolean function
or its representation by a specified structure, we denote by $Var(F)$
the set of variables of $F$. A truth assignment to $Var(F)$ on which $F$
is true is called a \emph{satisfying assignment} of $F$. A set $S$ of literals
represents the truth assignment to $Var(S)$ where variables occurring
positively in $S$ (i.e. whose literals in $S$ are positive) are assigned with $true$
and the variables occurring negatively are assigned with $false$.
We denote by $F_S$ a function whose set of satisfying assignments consists of $S'$
such that $S \cup S'$ is a satisfying assignment of $F$. We call $F_S$ a \emph{subfunction}
of $F$. 
%In other words, a Boolean function $F'$ is a subfunction of a Boolean function
%$F$ is $F'$ can be obtained from $F$ by giving a truth assignment to a subset of variables of $F$.

%In this paper we consider $3$ representations of Boolean functions: {\sc cnf}s, 
%Nondeterministic Read-Once Branching Programs ({\sc nrobp})s, and a restricted special
%case of the latter which we term  Normalized Free Binary Decision Diagrams ({\sc nfbdd}).
%\footnote{The {\sc fbdd} is just another name of the read-once branching program. We use
%it in order to avoid possible confusion caused by two similar abbreviations.} 
%Let us define the last two representations.

We define a Non-deterministic Read Once Branching Program ({\sc nrobp}) as a connected acyclic read-once \emph{switching-and-rectifier network} \cite{Yukna}.
That is, a {\sc nrobp} $Y$ implementing (realizing) a function $F$ is a directed acyclic graph (with possible multiple edges)
with one leaf, one root, and with some edges labelled by literals of the variables of $F$ in a way that 
there is no directed path having two edges labelled with literals of the same variable. We denote by $A(P)$ the set of literals
labeling edges of a directed path $P$ of $Y$.

The connection between $Y$ and $F$ is defined as follows.
Let $P$ be a path from the root to the leaf of $Y$. Then any extension of 
$A(P)$ to the truth assignment of all the variables of $F$ is a satisfying assignment of $F$.
Conversely, let $A$ be a satisfying assignment of $F$. Then there is a path $P$ from the root to
the leaf of $Y$ such that $A(P) \subseteq A$.

{\bf Remark.}
It is not hard to see that the traditional definition of {\sc nrobp} as a deterministic {\sc robp}
with guessing nodes \cite{RazFCT} can be thought as a special case of our definition (for any function that
is not constant $false$): remove from the former all the nodes from which the $true$ leaf is not reachable
and relabel each edge with the appropriate literal of the variable labelling its tail (if the original label
on the edge is $1$ then the literal is positive, otherwise, if the original label is $0$, the literal
is negative).

We say that a {\sc nrobp} $Y$ is \emph{uniform} if the following is true.
Let $a$ be a node of $Y$ and let $P_1$ and $P_2$ be $2$ paths from the root
of $Y$ to $a$. Then $Var(A(P_1))=Var((A(P_2))$. That is, these paths are labelled
by literals of the same set of variables. Also, if $P$ is a path from the root to the leaf of $Y$
then $Var(A(P))=Var(F)$. Thus there is a one-to-one correspondence between the sets of literals labelling
paths from the root to the leaf of $Y$ and the satisfying assignments of $F$.

{\bf All the {\sc nrobp}s considered in Sections 3-5 of this paper are uniform.}
This assumption does not affect our main result because
an arbitrary {\sc nrobp} can be transformed into a uniform one at the price of $O(n)$ times increase of
the number of edges. 
%We have not seen in the literature a transformation applying to the {\sc nrobp}
%as defined above. 
For the sake of completeness, we provide the transformation and its correctness
proof in the appendix. We use the construction described in the proof sketch of Proposition 2.1 of \cite{RaWiYao}.

For our counting argument we need a special case of {\sc nrobp} where \emph{all} the edges are labelled,
each node is of out-degree at most $2$ and $2$ out-edges of a node of degree exactly $2$ are labelled with
opposite literals of the same variable. We call this 
representation \emph{normalized free binary decision diagram} ({\sc nfbdd}).

%\footnote{{\sc fbdd} is another name of {\sc robp}
%widely usedin the area of verification and it is not hard to see that {\sc
%It is easy to show that any function $F$ that is not constant $false$
%can be represented by a {\sc nfbdd}.}

We need additional terminology regarding {\sc nfbdd}. 
We say that each non-leaf node $a$ is \emph{labelled} by the variable whose literals
label its out-edges and denote this variable by $Var(a)$.
Further on, we refer to the
out-going edges of $a$ labeled by, respectively, positive and negative literals
of $Var(a)$ as \emph{positive} and \emph{negative} out-going edges of $a$. The heads of these edges
are respective \emph{positive} and \emph{negative} out-neighbours of $a$ (if both edges have the same
head then these out-neighbours coincide). Note that given a labelling on nodes, there will be no 
loss of information if all the positive edges are labelled with $1$ and all the negative edges are labelled
with $0$: the information about the labelling variable can be read from the tail of each edge and hence only
the information about the sign of the labelling literal is needed. It follows that, for instance an {\sc obdd}
with all the nodes from which the $yes$-leaf cannot be reached being removed is, in essence, an {\sc nfbdd}.
Consequently, any Boolean function that is not constant \emph{false} can be represented by an {\sc nfbdd}. 

\begin{figure}[h]
\includegraphics[height=5cm]{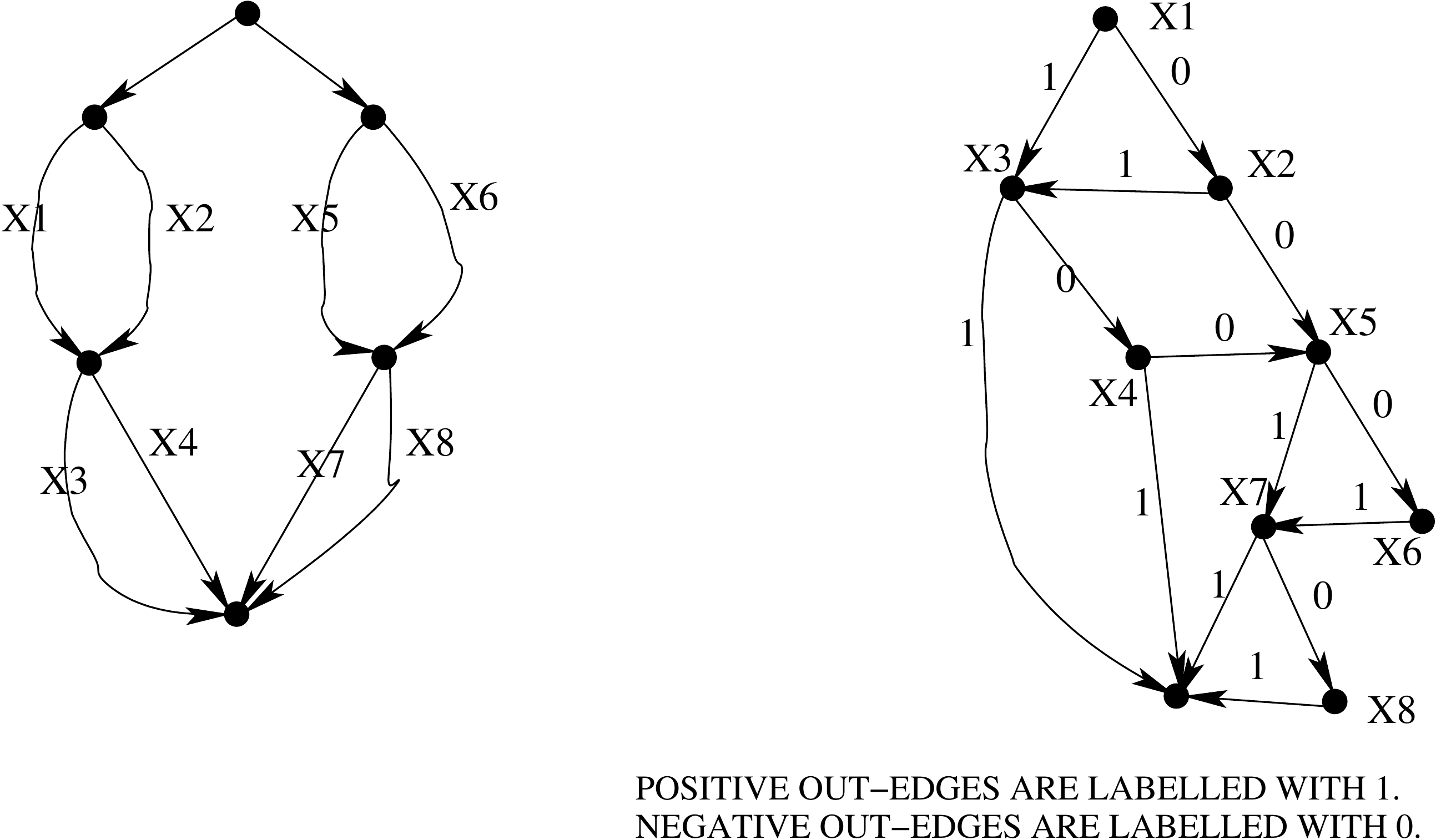}
\caption{{\sc nrobp} and {\sc nfbdd} for function $((x_1 \vee x_2)\wedge(x_3 \vee x_4)) \vee ((x_5 \vee x_6) \wedge (x_7 \vee x_8))$}
\label{models}
\end{figure}

Figure \ref{models} illustrates a {\sc nrobp} and a {\sc nfbdd} for a particular function.

Given a graph $G$, its \emph{tree decomposition} is a pair $(T,{\bf B})$ where $T$ 
is a tree and ${\bf B}$ is a set of bags $B(t)$ corresponding to the vertices $t$ of $T$.
Each $B(t)$ is a subset of $V(G)$ and the bags obey the rules of \emph{union} (that is, $\bigcup_{t \in V(T)} B(t)=V(G)$),
\emph{containment} (that is, for each $\{u,v\} \in E(G)$ there is $t \in V(t)$ such that $\{u,v\} \subseteq B(t)$),
and \emph{connectedness} (that is for each $u \in V(G)$, the set of all $t$ such that $u \in B(t)$ induces a subtree of $T$).
The \emph{width} of $(T,{\bf B})$ is the size of the largest bag minus one. The treewidth of $G$ is the smallest width of a tree
decomposition of $G$.

Given a {\sc cnf} $\phi$, its \emph{primal graph} has the set of vertices corresponding to the variables of $\phi$.
Two vertices are adjacent if and only if there is a clause of $\phi$ where the
corresponding variables both occur.
\section{The main result}
A \emph{monotone} 2-{\sc cnf}s has clauses
of the form $(x \vee y)$ where $x$ and $y$ are two distinct variables. 
Such {\sc cnf}s can be put in one-to-one correspondence
with graphs that do not have isolated vertices. In particular, let $G$ be such
a graph. Then $G$ corresponds to a 2{\sc cnf} $\phi(G)$ whose set of variables
is $\{x_v|v \in V(G)\}$ and the set of clauses is $\{(x_u \vee x_v)|\{u,v\} \in E(G)\}$.
It is not hard to see that $G$ is a primal graph of $\phi(G)$, hence we can refer to the 
treewidth of $G$ as the the primal graph treewidth of $\phi(G)$.
For $u \in V(G)$, denote by $Var(u)$ the variable of $\phi(G)$ corresponding to 
$u$ and for $V' \subseteq V(G)$, let $Var(V')=\{Var(u)|u \in V'\}$.
Conversely, let $x$ be a literal of a variable of $\phi(G)$. Then the corresponding
vertex of $G$ is denoted by $Vert(x)$. If $X'$ is a set of literals of variables of $\phi(G)$
then $Vert(X')=\{Vert(x)|x \in X'\}$. 

The following theorem is the main result of this paper. 

\begin{theorem} \label{maintheor}
There is a constant $c$ such that
for each $k \geq 50$ there is an infinite class ${\bf G}$ of graphs each
of treewidth of at most $k$ such that for each $G \in {\bf G}$,
the smallest {\sc nrobp} equivalent to $\phi(G)$ is of size at least ${n}^{k/c}$,
where $n$ is the number of variables of $\phi(G)$.
\end{theorem}

In order to prove Theorem \ref{maintheor}, we 
introduce the notion of \emph{distant matching width} ({\sc dmw})
of a graph and state two theorems proved in the subsequent two sections.
One claims that if the max-degree of $G$ is bounded then the size of a {\sc nrobp} realizing $\phi(G)$
is exponential in the {\sc dmw} of $G$. The other theorem claims that for each sufficiently large $k$ 
there is an infinite class of graphs of a bounded degree and of treewidth at most $k$ whose {\sc dmw} 
is at least $b*log n*k$ for some universal constant $b$. 
Theorem \ref{maintheor} will follow as an immediate corollary of these two theorems. 

\begin{definition}{\bf Matching width.}\\
Let $SV$ be a \emph{permutation} of $V(G)$ of vertices of a graph. 
and et $S_1$ be a \emph{prefix} of $SV$ (i.e. all vertices of $SV \setminus S_1$ are ordered after
$S_1$). The \emph{matching width} of $S_1$ is the size of the largest matching
consisting of the edges between $S_1$ and $V(G) \setminus S_1$
(we sometimes treat sequences as sets, the correct use will be always clear
from the context). The \emph{matching width} of $SV$ is the largest \emph{matching
width} of a prefix of $SV$. The \emph{matching width} of $G$, denoted by $mw(G)$, is the smallest 
matching width of a permutation of $V(G)$. 
\end{definition}

\paragraph{Remark.} The above definition of matching width is a special 
case of a more general notion of \emph{maximum matching width} as defined
in \cite{VaThesis}. In particular, our notion of matching width can be seen
as a variant of maximum matching width of \cite{VaThesis} 
where the tree $T$ involved in the definition is a caterpillar.
Also, \cite{VaThesis} considers the notion of maximum induced matching width requiring
that that the ends of different edges of the witnessing matching are not adjacent.
We need to impose a stronger constraint on the witnessing matching as specified below. 

\begin{definition} {\bf Distant matching}\\
A matching $M$ of $G$ is \emph{distant} 
if it is induced (no neighbours between vertices incident to distinct edges of $M$)
and also no two vertices incident to distinct edges of $M$ have a common neighbor.
\end{definition}

\begin{definition} {\bf Distant matching width}\\
Distant matching width ({\sc dmw}) is defined analogously to matching width with
`matching' replaced by `distant matching'.
The {\sc dmw} of graph $G$ is denoted by $dmw(G)$.
Put it differently, $dmw(G)$ equals the largest $t$ such that for any permutation
$SV$ of $V(G)$ there is a partition $V_1,V_2$ into a prefix and a suffix such that 
there is a \emph{distant} matching of size $t$ consisting of edges with one end in $V_1$ and the other end
in $V_2$. 
\end{definition}

To illustrate the above notions recall that $C_n$ and $K_n$ respectively
denote a cycle and a complete graph of $n$ vertices. Then, for a sufficiently large $n$,
$mw(C_n)=dmw(C_n)=2$. On the other hand $mw(K_n)=\lfloor n/2 \rfloor$ while $dmw(K_n)=1$. 

\begin{theorem} \label{nrobplbdmw}
For each integer $i$ there is a constant $a_i$ such that for any graph $G$ the size of
{\sc nrobp} realizing $\phi(G)$ is at least $2^{dmw(G)/a_x}$
where $x$ is the max-degree of $G$. 
\end{theorem}

%The proof of Theorem \ref{nrobplvdmw} is provided in Section \ref{lbmainproof}.

\begin{theorem} \label{dmwtw}
There is a constant $b$ such that
for each $k \geq 50$ there is an infinite class ${\bf G}$ of graphs of degree at most $5$ 
such that the treewidth of all the graphs of $G$ is at most $k$ and for each $G \in {\bf G}$ 
the matching width is at least $(log n*k)/b$ where $n=|V(G)|$.
\end{theorem}

%The proof of Theorem \ref{dwtw} is provided in Section \ref{dmwmainproof}.

Now we are ready to prove Theorem \ref{maintheor}.

{\bf Proof of Theorem \ref{maintheor}.}
Let ${\bf G}$ be the class whose existence is claimed by Theorem \ref{maintheor}.
By theorem \ref{nrobplbdmw}, for each $G \in {\bf G}$ the size of a {\sc nrobp} 
realizing $\phi(G)$ is of size at least $2^{dmw(G)/a_5}$.
Further on, by Theorem \ref{dmwtw}, $dwm(G) \geq (log n*k)/b$, for some constant $b$
Substituting the inequality for $dmw(G)$ into $2^{dmw(G)/a_5}$, we get that the size
of a {\sc nrobp} is at least $2^{log n *k/c}$ where $c=a_5*b$. Replacing $2^{log n}$ by $n$
gives us the desired lower bound. $\blacksquare$

From now on, the proof is split into two independent parts: Section \ref{lbmainproof} proves
Theorem \ref{nrobplbdmw} and Section \ref{dmwmainproof} proves Theorem \ref{dmwtw}.

\section{Proof of Theorem \ref{nrobplbdmw}} \label{lbmainproof}
Let $S$ be an assignment to a subset of variables of $\phi(G)$ and $V' \subseteq V(G)$.
We say that $V'$ \emph{covers} $S$ if all the variables of $Var(V')$ occur positively in $S$.
%provide an example 
Furthermore, we call $V' \subseteq V(G)$ a \emph{distant independent set} {\sc dis} of $G$ if 
$V'$ is an independent set of $G$ and, in addition, no two vertices of $V'$ have a 
common neighbour.

In order to prove Theorem \ref{nrobplbdmw}, we first introduce Lemma  \ref{coverlb} 
(proved in Section \ref{coverbound}) stating that at least $2^{t/a}$ {\sc dis}es are needed to cover
all the satisfying assignments of $\phi(G)$ where $a$ is a constant depending on the max-degree
of $G$. After that we show that a {\sc nrobp} $Z$ of $\phi(G)$ always has a root-leaf (node)
cut $K$ such that each node $u$ of the cut can be associated with a {\sc dis} of size $dmw(G)$ such that
for all the root-leaf paths $P$ passing through $u$, $A(P)$ (recall that $A(P)$ is the set of labels on the edges 
of $P$) is covered by this {\sc dis}.
Since each satisfying assignment of $\phi(G)$  is $A(P)$ of some root-leaf path $P$ and since $P$
passes through a node of $K$ (due to $K$ being a root-leaf cut of $Z$), we conclude that all the satisfying
assignments of $\phi(G)$ are covered by the considered family of {\sc dis}es. Using Lemma \ref{coverlb},
we will conclude that this set of {\sc dis}es is large and hence the set $K$ and, consequently, $Z$ are large as well.

\begin{lemma} \label{coverlb}
For each $i$ there is a constant $a_i$ such that for any $t$ the number of {\sc dis}es of $G$ of size 
$t$ needed to cover all the satisfying assignments of $\phi(G)$ is at least $2^{t/a_x}$ where
$x$ is the max-degree of $G$. Put it differently, if ${\bf M}$ is a family of {\sc dis}es of size $t$
such that each satisfying assignment of $\phi(G)$ is covered by at least one element of ${\bf M}$ then 
$|{\bf M}| \geq 2^{t/a_x}$
\end{lemma}

\begin{lemma} \label{dmw2}
Let $P$ be a path from the root to the leaf of a {\sc nrobp} realizing $\phi(G)$.
Then $P$ has a node $u(P)$ for which the following holds. Let $P_1$ be the prefix
of $P$ ending at $u(P)$ and let $P_2$ be the suffix of $P$ starting at $u(P)$.
Denote  $Vert(A(P_1))$ and $Vert(A(P_2))$ by $V_1$ and $V_2$, respectively.
Then $G$ has a distant matching $M$ of size $dmw(G)$ such that one end of each edge of $M$
is in $V_1$ and the other end is in $V_2$.
\end{lemma}

{\bf Proof.}
%Essentially follows from the definition of {sc dmw}.
%The sequence of labels of any path of $Z$ from the root to the leaf corresponds
%to a permutation of vertices of $G$. By definition of {sc dmw}, it is always possible
%to find a partition into a prefix and suffix so that the edges of $G$ between
%the partition classes witness $dmw(G)$. Back to the paths of $Z$, we simply take
%the path 
Let $SL$ be the sequence of $Var(A(P))$ listed by the chronological
order of the occurrence of respective literals on the edges of $P$ being explored from the root to the leaf.
Due to the uniformity and the read-onceness of $Z$, $SL$ is just a permutation of the variables of $\phi(G)$.
By definition of $\phi(G)$, $SL$ corresponds to a permutation $SV$ of $V(G)$. 
Moreover, for a prefix $V_1$ of $SV$, there is a
partition of $P$ into a prefix $P_1$ and a suffix $P_2$ such that $V_1=Vert(A(P_1))$ and 
$V\setminus V_1=Vert(A(P_2))$. Indeed, take a prefix $P_1$ including precisely the first $|V_1|$ labels
by letting the final node of $P_1$ to be the head of the edge carrying the $|V_1|$-th label.
If the desired equalities are not satisfied then the vertices of $SV$ are listed in an order different from
the order of occurrence of the corresponding variables in $SL$, a contradiction.
It remains to recall that by definition of $dmw(G)$, a witnessing partition $V_1,V_2$ exists for any
permutation $SV$ of $V(G)$ and to take the prefix and a suffix of $P$ corresponding to such $V_1$ and $V_2$.
$\blacksquare$.  

The cut we will consider for the purpose of proving Theorem \ref{nrobplbdmw} will be the set of nodes $u(P)$ for all
the paths $P$ of $Z$ from the root to the leaf. The next lemma will allow us to transform the matching associated
with each vertex of this cut into a {\sc dis} by taking one vertex of each edge of this matching.

\begin{lemma} \label{dmw3}
Let $Z$ be a {\sc nrobp} realizing $\phi(G)$ of a graph $G$.
Let $P$ be a path from the root to the leaf of $Z$ and let $a$ be a vertex of this path.
Let $P_1$ be the prefix of $P$ ending at $a$ and let $P_2$ be the suffix of $P$ beginning at $a$.
Denote $Vert(A(P_1))$ and $Vert(A(P_2))$ by $V_1$ and $V_2$ respectively. Let $\{v_1,v_2\}$ be an
edge of $G$ such that $v_1 \in V_1$ and $v_2 \in V_2$. Then either $\{v_1\}$ covers all the assignments
$A(P')$ such that $P'$ is a root-leaf path of $Z$ passing through $a$ or this is true regarding $\{v_2\}$. 
\end{lemma} 

{\bf Proof.}
Let $x_1,x_2$ be the respective variables of $\phi(G)$ corresponding to $v_1$ and $v_2$.
Recall that by definition, $\phi(G)$ contains a clause $(x_1 \vee x_2)$. Suppose that the statement
of the lemma is not true. That is, there are 2 paths $P'$ and $P''$ from the root to the leaf of $Z$,
both passing through $a$ and such that $P'$ is not covered by $v_1$ and $P''$ is not covered by $v_2$.
Let $P'_1,P'_2$ be the prefix and suffix of $P'$ with $a$ being the final vertex of $P'_1$ and the initial
vertex of $P'_2$. Let $P''_1$ and $P''_2$ be the analogous partition of $P''$. 

Observe that due to the uniformity of $Z$, $Vert(A(P_1))=Vert(A(P'_1))$. In particular, $v_1 \in Vert(A(P'_1))$
and hence the occurrence of $x_1$ in $A(P')$, in fact belongs to $A(P'_1)$. Since $\{v_1\}$ does not cover $P'$,
$\neg x_1 \in A(P')$ and hence $\neg x_1 \in A(P'_1))$. Analogously, $Ver(A(P_1))=Vert(A(P''_1))$ and hence
$Vert(A(P''_1))$ does not contain $v_2$ leading to the conclusion that $\neg x_2 \in A(P''_2)$.
By construction, $P'_1 \cup P''_2$ is a path from the root to the leaf of $Z$ and hence $A(P'_1 \cup P''_2)=
A(P'_1)\cup A(P''_2)$ is a satisfying assignment of $\phi(G)$. However, this is a contradiction since  
$A(P'_1)\cup A(P''_2)$ contains $\{\neg x_1,\neg x_2\}$ falsifying a clause of $\phi(G)$. $\blacksquare$

Now we are ready to prove Theorem \ref{nrobplbdmw}.

{\bf Proof of Theorem \ref{nrobplbdmw}.}
For each path $P$ from the root to the leaf of $Z$, pick a vertex $u(P)$ as specified in
Lemma \ref{dmw2}. Let $\{u_1, \dots, u_q\}$ be the set of all such $u(P)$. By construction each of them is neither 
the root nor the leaf and each path from the root to the leaf passes through some $u_i$.
So, they indeed constitute a root-leaf cut of $Z$. 
Further on, for each $u_i$ specify a witnessing path $P^i$ such that $u_i=u(P^i)$ and such that
$P^i_1$ is the prefix of $P^i$ ending at $u_i$ and $P^i_2$ is the suffix of $P^i$ beginning at $u_i$.
By definition of $u(P_i)$, there is distant matching $M_i$ of size $dmw(G)$ such that one end of
each edge of $M_i$ belongs to $Vert(A(P^i_1))$ and the other end belongs to $Vert(A(P^i_2))$.
By Lemma \ref{dmw3} we can choose one end of each edge of $M_i$ that covers $A(P')$ for all $P'$ passing
through $u_i$. Let $B_i$ be the set of the chosen ends. By definition of a distant matching these
vertices are mutually non-adjacent and do not have common neighbours. It follows that each $B_i$ is a {\sc dis} 
of $G$ of size $dmw(G)$. Moreover, by construction, each $B_i$ covers $A(P')$ for all $P'$ passing through $u_i$.
It follows that each satisfying assignment $A'$ of $\phi(G)$ is covered by some $B_i$.
Indeed, by definition of {\sc nrobp}, $Z$ has a path $P'$ from the root to the leaf such that $A(P')=A'$.
Since $\{u_1, \dots, u_q\}$ is a root-leaf cut of $Z$, $P'$ passes through some $u_i$. Consequently, $A'=A(P')$ is covered
by $B_i$. It follows from Lemma \ref{coverlb} that $q \geq 2^{dmw(G)/a_x}$ where 
$x$ is the max-degree of $G$, confirming the theorem. %would be good to reformulate
$\blacksquare$. 

\subsection{Proof of Lemma \ref{coverlb}} \label{coverbound}
In order to prove Lemma \ref{coverlb}, we assume that $\phi(G)$ is represented as a {\sc nfbdd} $Y$.
%Then we use a weighted counting approach inspired by a probabilistic argument as in e.g. \cite{RaWiYao}.
\begin{comment}
The proof strategy is similar to the probabilistic argument as in e.g. \cite{RaWiYao}.
Roughly speaking, we consider a random walk moving from the root to the leaf of $Y$, on each node randomly
choosing an out-going edge. Such a walk eventually reaches a leaf with probability $1$, while we show that the probability of
the corresponding assignment being covered by the considered set $B$ is exponentially small in the size of $B$. Hence the number
of such sets $B$ must be large. 
For the technical description, we present the above probabilistic argument in a form of weighted counting. 
\end{comment}
For each edge $e$ of $Y$ we assign weight
$w(e)$ as follows. For a vertex $a$ of $Y$ with $2$ leaving edges, the weight of each edge is $0.5$. 
If $a$ has only one leaving edge, the weight of this edge is $1$. The weight $w(P)$ of a path $P$ of $Y$ is defined as follows.
If $P$ consists of a single vertex then $w(P)=1$. Otherwise $w(P)$ is the product of weights of its edges. Let ${\bf P}$ be a set
of paths. Then $w({\bf P})=\sum_{P \in {\bf P}} w(P)$ defines the weight of ${\bf P}$.
The following proposition immediately follows from the non-negativity of weights.

\begin{proposition} \label{ubound}
Let ${\bf P}'_1, \dots, {\bf P}'_x$ be a sets of paths of $Y$.
Then $w(\bigcup_{i=1}^x {\bf P}'_i) \leq \sum_{i=1}^x w({\bf P}'_i)$  
\end{proposition}

Let $a$ be a node of $Y$ and let ${\bf P}_a$ be the set of all paths from $a$
to the leaf of $Y$. Then the following can be easily noticed. 

\begin{proposition} \label{onebound}
$w({\bf P}_a)=1$. 
\end{proposition}

Let $S \subseteq V(G)$. Let ${\bf P}_a^S$ be the subset of ${\bf P}_a$ consisting of
all $P$ such that $A(P)$ is covered by $S$. We will show 
that if $S$ is a {\sc dis} of $G$ and $G$ is of bounded degree then $w({\bf P}_{rt}^S)$
is exponentially small in $|S|$ where $rt$ is the root of $Y$. Then we will note that if $S_1, \dots, S_q$ are {\sc dis}es such that each satisfying
assignment is covered by one of them then ${\bf P}_{rt}^{S_1} \cup \dots \cup {\bf P}_{rt}^{S_q}={\bf P}_{rt}$
and hence $w({\bf P}_{rt}^{S_1})+ \dots w({\bf P}_{rt}^{S_q}) \geq w({\bf P}_{rt}^{S_1} \cup \dots \cup {\bf P}_{rt}^{S_q})=1$.
Consequently, $q$ must be exponentially large in $|S|$, implying the lemma. 
This weighted counting approach is inspired by a probabilistic argument as in e.g. \cite{RaWiYao}.

\begin{comment}
We are now going to state an upper bound on $w({\bf P}_a^S)$
for a case where $S$ is a specially defined {\sc dis}. A special case of this statement is 
that if $B$ is an arbitrary {\sc dis} of $G$ and $G$ is of bounded degree
$w({\bf P}_{rt}^B)$ is exponentially small is $|B|$ where $rt$ is the root of $Y$.

It will immediately follow
that in order  will imply Lemma \ref{coverlb}.
Before introducing the statement, we need to

Let us first extend our terminology by introducing some notions related 
to the nodes of $Y$. 
\end{comment}

\begin{comment}
For a node $a$ of $Y$, and a path $P$ from the root of $Y$ to $a$, we denote
$V(G) \setminus Vert(A(P))$ by $Vert_a$. Notice that due to the uniformity of $Y$,
$V(G) \setminus Vert(A(P))$ does not depend on a particular choice of $P$ as long as it is 
a path from the root of $Y$ to $a$. Hence $Vert_a$ is well defined and means the set of vertices of $G$ corresponding
to the variables that have not been assigned by a path from the root to $a$.
\end{comment}
We denote by $Vert_a$ the set of vertices of $G$ corresponding
to the variables that have not been assigned by a path from the root to $a$.

We denote by $Free_a$ the subset of $Vert_a$ consisting of all vertices $v$ such that there is a path
$P \in {\bf P}_a$ with $\neg Var(v) \in A(P)$. (This is only possible if no label $\neg Var(u)$ occurs on a path
from the root to $a$ such that $u$ is a neighbour of $v$. That is, $v$ is `free'in the sense that it is not constrained
by such an occurrence.)   

For $v \in V(G)$ we denote by $ld_a(v)$ the number of neighbours of $v$ in $Vert_a$ (`ld' stands for `local degree').

For $B \subseteq V(G)$, we define $rw_a(B)$ as follows (`rw' stands for `relative weight').
If $B=\emptyset$ then $rw_a(B)=1$. Otherwise, let $v \in B$. Then $rw_a(B)=(1-2^{-(ld_a(v)+1)})*rw_a(B \setminus \{v\})$.
For a non-empty $B$, $rw_a(B)$ can be seen as
$\prod_{v \in B} (1-2^{-(ld_a(v)+1)})$.

\begin{comment}
The next lemma (Lemma \ref{deepcover}) states the $w({\bf P}^B_a)$ is exponentially small in the size of $B$.
After that we will show how this lemma implies Lemma \ref{coverlb}. In the rest of the section we provide a
proof of Lemma \ref{deepcover}.
\end{comment}

The following is our main technical argument.

\begin{lemma}\label{deepcover}
Let $a$ be a node of $Y$ and let $B \subseteq Free_a$ be a {\sc dis} of $G$.
Then $w({\bf P}_a^B) \leq rw_a(B)$. 
\end{lemma}

In the rest of the section we prove Lemma \ref{coverlb} and then provide a proof of Lemma \ref{deepcover}.

{\bf Proof of Lemma \ref{coverlb}.}
Denote the root of $Y$ by $rt$. It is not hard to see that $Free_{rt}=V(G)$ (for each variable of $\phi(G)$
there is a satisfying assignment where this variable appears negatively), hence Lemma \ref{deepcover}
applies to ${\bf P}^B_{rt}$ for an arbitrary {\sc dis} $B$ of $G$. Moreover, $ld_{rt}(v)$ 
is simply $d(v)$, the degree of $v$ in $G$. Therefore, it follows from Lemma \ref{deepcover} that
$w({\bf P}_{rt}^B) \leq \prod_{v \in B}(1-2^{-(d(v)+1)})$. Since $d(v) \leq x$ (recall that $x$ denotes the max-degree of $G$), 
$\prod_{v \in B}(1-2^{-(d(v)+1)}) \leq \prod_{v \in B}(1-2^{-(x+1)})=(1-2^{-(x+1)})^t$ where $t=|B|$.
Let $B_1, \dots, B_q$ be {\sc dis}es of size $t$ that cover all the satisfying assignments of 
$\phi(G)$. It follows that ${\bf P}^{B_1}_{rt} \cup \dots \cup {\bf P}^{B_q}_{rt}={\bf P}_{rt}$.
Indeed, the left-hand side is contained in the right-hand side by definition, so let $P \in {\bf P}_{rt}$.
Then, by definition, of $Y$, $A(P)$ is a satisfying assignment of $\phi(G)$. By definition of $B_1, \dots, B_q$,
there is some $B_i$ covering $A(P)$. Then it follows that $P \in {\bf P}^{B_i}_{rt}$. 
Combining propositions \ref{ubound} and \ref{onebound}, we obtain:
%\begin{align*}
$1=w({\bf P}_{rt})=w(\bigcup_{i=1}^q {\bf P}^{B_i}_{rt}) \leq 
\sum_{i=1}^q w({\bf P}^{B_i}_{rt}) \leq q*(1-2^{-(x+1)})^t$
%\end{align*}

It follows that $q \geq (\frac{1}{1-2^{-(x+1)}})^t$.
Clearly, for each $x$ there is a constant $a_x$ such that 
$\frac{1}{1-2^{-(x+1)}}$ can be represented as $2^{1/a_x}$.
Hence the bound $q \geq 2^{t/a_x}$ follows. $\blacksquare$

To prove Lemma \ref{deepcover}, we need a number of auxiliary statements 
provided below.

\begin{lemma} \label{onepos}
Let $a$ be a non-leaf node
of $Y$ having only one out-neighbour. Then this out-neighbour is positive.
\end{lemma}

\begin{lemma} \label{freeaprime}
Let $a$ be a node of $Y$ and let $a'$ be an out-neighbour of $a$.
Denote $Vert(Var(a))$ by $v$ and let $B \subseteq Free_a$. Then the following statements hold.
\begin{itemize}
\item If $v \in B$ then $B \setminus \{v\} \subseteq Free_{a'}$.
\item If there is $w \in B$ such that $\{v,w\} \in E(G)$ and $a'$ is a negative out-neighbour of $a$ then
$B \setminus \{w\} \subseteq Free_{a'}$.
\item In all other cases, $B \subseteq Free_{a'}$. 
\end{itemize}
\end{lemma}

Let $(a,a')$ be an edge of $Y$ and let $P$ be a path of $Y$ starting at $a'$
Then $(a,a')+P$ denotes the path obtained by concatenating $(a,a')$ and $P$.
Let ${\bf P}$ be a set of paths all starting at $a'$.
Then $(a,a')+{\bf P}=\{(a,a')+P|P \in {\bf P}\}$.

\begin{proposition} \label{pathincrease}
$w((a,a')+{\bf P})=w(a,a')*w({\bf P})$.
\end{proposition}

\begin{lemma} \label{pathdecomp}
Let $a$ be a node of $Y$.
Denote $Vert(Var(a))$ by $v$. Let $B$ be a {\sc dis} of $G$. Then the following
statements are true.
\begin{itemize}
\item Assume that $v \in B$. Then ${\bf P}^B_a \subseteq (a,a')+{\bf P}^{B \setminus\{v\}}_{a'}$ where $a'$
is the positive out-neighbour of $a$.
\item Otherwise, ${\bf P}^B_a \subseteq \bigcup_{a' \in N^+_Y(a)} ((a,a')+{\bf P}^B_{a'})$,
      where $N^+_Y(a)$ is the set of out-neighbours of $a$.
\end{itemize}
\end{lemma}

%$\prod_{v \in B} (1-2^{-(ld_a(v)+1)})$ by $rw_a(B)$
\begin{lemma} \label{rwdecomp}
Let $a$ be a node of $Y$, let $a'$ be an out-neighbour of $a$, and let $B$ be a {\sc dis} of $G$.
Denote $Vert(Var(a))$ by $v$. Then the following statements hold.
\begin{itemize}
\item Assume that $v \in B$. Then $rw_{a'}(B \setminus \{v\})=rw_a(B)/(1-2^{-(ld_a(v)+1)})$.
\item Assume there is $w \in B$ such that $\{v,w\} \in E(G)$. Then 
$rw_{a'}(B \setminus \{w\})=rw_a(B)/(1-2^{-(ld_a(w)+1)})$ and
$rw_{a'}(B)=rw_a(B)*\frac{1-2^{-ld_a(w)}}{(1-2^{-(ld_a(w)+1)})}$.
\item If none of the above assumptions is true then 
$rw_{a'}(B)=rw_a(B)$.
\end{itemize}
\end{lemma}

{\bf Proof of Lemma \ref{deepcover}.}
The proof is by induction on the reverse topological ordering of the nodes of $Y$ (leaves first
and if a non-leaf node is considered, the lemma is assumed correct for all its out-neighbours).
Let $a$ be a leaf of $Y$. Clearly, $Free_a=\emptyset$ and hence we can only consider the set
${\bf P}^{\emptyset}_a$ consisting of a single path including node $a$ itself. It follows that
$w(P^{\emptyset}_a)=1$. On the other hand, $rw_a(\emptyset)=1$ by definition. Hence the lemma holds
in the considered case.

Assume now that $a$ is not a leaf and denote $Vert(Var(a))$ by $v$.

{\bf Suppose first that $v \in B$.} Since $v \in Free_a$, there is a path $P^* \in {\bf P}_a$
such that $Var(v)$ occurs negatively in $P^*$. That is $P^*$ contains a node $a^*$ such that
$Var(a^*)=Var(v)$ and the leaving edge of $a^*$ included in $P^*$ is the negative one.
Due to the read-onceness, the only node of $P^*$ whose associated variable is $Var(v)$ is $a$.
Consequently, $a$ has a a leaving negative edge. It follows from Lemma \ref{onepos} that 
$a$ has $2$ out-neighbours and hence the weight of each leaving edge is $0.5$.

Let $a'$ be the positive out-neighbour of $a$. 
Combining Lemma \ref{pathdecomp} and Proposition \ref{pathincrease}, we obtain,
$w({\bf P}_a^B) \leq w((a,a')+{\bf P}_{a'}^{B \setminus \{v\}})=w(a,a')*w({\bf P}_{a'}^{B \setminus \{v\}})=
 0.5*w({\bf P}_{a'}^{B \setminus \{v\}})$. 

By Lemma \ref{freeaprime}, $B \setminus \{v\} \subseteq Free_{a'}$. 
By the induction assumption and Lemma \ref{rwdecomp}, 
$w({\bf P}_{a'}^{B \setminus \{v\}}) \leq rw_{a'}(B \setminus \{v\})=rw_a(B)/(1-2^{-(ld_a(v)+1)})$.
It follows that $w({\bf P}_a^B) \leq 0.5* rw_a(B)/(1-2^{-(ld_a(v)+1)})$. Since $1-2^{-(ld_a(v)+1)} \geq 0.5$,
$w({\bf P}_a^B) \leq rw_a(B)$.

{\bf Suppose that $v$ is a neighbour of some $w \in B$}. 
Assume first that $a$ has only one out-neighbour $a'$. According to Lemma \ref{onepos}, $a'$ is a positive 
out-neighbour. Combining Lemma \ref{pathdecomp}, Proposition \ref{pathincrease}, and taking into account that
$w(a,a')=1$, we obtain the following.
$w({\bf P}_a^B) \leq w((a,a')+{\bf P}_{a'}^{B})=w({\bf P}_{a'}^{B})$. By Lemma \ref{freeaprime}, $B \subseteq Free_{a'}$.
By the induction assumption combined with Lemma \ref{rwdecomp}, we obtain: 
$w({\bf P}_a^B) \leq w({\bf P}_{a'}^{B}) \leq rw_{a'}(B)=rw_a(B)*\frac{1-2^{-ld_a(w)}}{(1-2^{-(ld_a(w)+1)})}$.
The numerator in the last item is smaller than the denominator and hence $w({\bf P}_a^B) \leq rw_a(B)$ follows. 

Assume now that in addition to $a'$, $a$ has the negative out-neighbour $a''$.
According to Lemma \ref{pathdecomp},
${\bf P}_a^B \subseteq ((a,a')+{\bf P}_{a'}^B) \cup ((a,a'')+{\bf P}_{a''}^B)$. 
Since any assignment covered by $B$ is also covered by a subset of $B$,
${\bf P}_{a''}^B \subseteq {\bf P}_{a''}^{B \setminus \{w\}}$ and hence
${\bf P}_a^B \subseteq ((a,a')+{\bf P}_{a'}^B) \cup ((a,a'')+{\bf P}_{a''}^{B \setminus \{w\}})$.  
Note that $B \subseteq Free_{a'}$ and $B \setminus \{w\} \subseteq Free_{a''}$ by Lemma \ref{freeaprime}.
Combining the induction assumption with Proposition \ref{ubound}, with Lemma \ref{rwdecomp}, and with the fact
that $w((a,a'))=w((a,a''))=0.5$, we obtain,
${\bf P}_a^B \leq 0.5*rw_a(B)*\frac{1-2^{-ld_a(w)}}{(1-2^{-(ld_a(w)+1)})}+0.5*rw_a(B)/(1-2^{-(ld_a(w)+1)})=
0.5rw_a(B)\frac{2-2^{-ld_a(w)}}{(1-2^{-(ld_a(w)+1)})}=0.5*2*rw_a(B)\frac{1-2^{-(ld_a(w)+1)}}{(1-2^{-(ld_a(w)+1)})}=rw_a(B)$.

{\bf Suppose that none of the previous assumptions occur.}
By Corollary \ref{freeaprime}, $B \subseteq Free_{a'}$ for any out-neighbour of $a$.
By the induction assumption, combined with Lemma \ref{rwdecomp}, ${\bf P}^B_{a'} \leq rw_{a}^B$.
Hence, by Lemma \ref{pathdecomp} combined with Proposition \ref{pathincrease} and Proposition \ref{ubound},
we obtain, 
$w({\bf P}_a^B) \leq \sum_{a' \in N^+_Y(a)} w(a,a')*rw_a(B)=rw_a(B)$. $\blacksquare$

\section{Proof of Theorem \ref{dmwtw}} \label{dmwmainproof}
\begin{comment}
We start our reasoning from a statement that for graphs of bounded degree a lower 
bound on the \emph{matching width} of $G$ implies a lower bound on the \emph{distant matching
width}.

\begin{proposition} \label{dmw1}
For a graph $G$ with maxdegree $c$, $dmw(G) \geq mw(G)/(2c^2+2c+1)$. 
%Recall that $dmw(G)$ denotes the {\sc dmw} of $G$ and $mw(G)$ denotes the matching width of $G$.
\end{proposition}

It follows from Proposition \ref{dmw1} that to prove Theorem \ref{dmwtw}, it is sufficient
to introduce a class of graphs of bounded degree  with the required upper bound on the treewidth
and the required lower bound on the \emph{matching width}. 
\end{comment}

\begin{figure}[h]
\includegraphics[height=2.5cm]{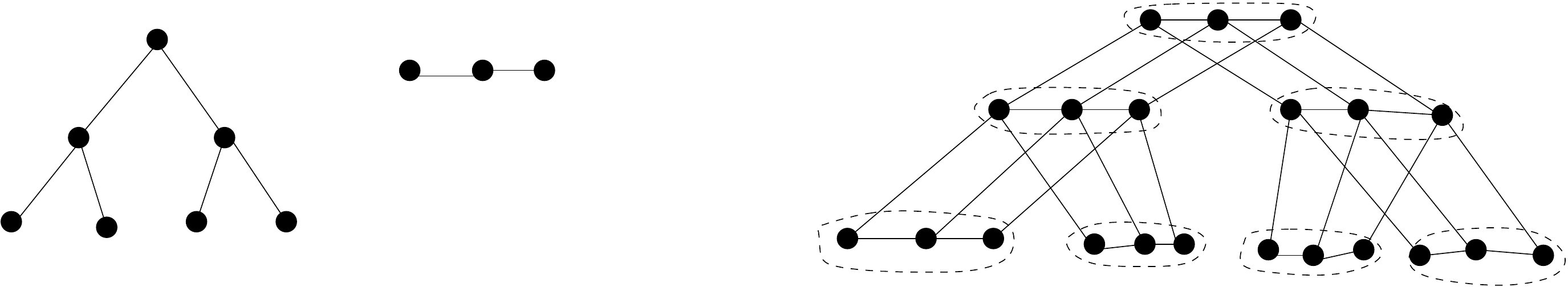}
\caption{Graphs from the left to the right: $T_3,P_3,T_3(P_3)$. The dotted ovals surround the copies of $P_3$ in $T_3(P_3)$.}
\label{trees}
\end{figure}

In order to prove Theorem \ref{dmwtw}, we consider graphs
$T(H)$ where $T$ is a tree and $H$ is an arbitrary graph. 
Then $T(H)$ is a graph having disjoint copies of $H$ in one-to-one correspondence with the vertices of $T$. 
For each pair $t_1,t_2$ of adjacent vertices of $T$, the corresponding copies are
connected by making adjacent the pairs of \emph{same} vertices of these copies.
Put it differently, we can consider $H$ as a labelled graph where all vertices are associated with distinct
labels. Then for each edge $\{t_1,t_2\}$ of $T$, edges are introduced between the vertices of the corresponding copies having
the same label. An example of this construction is shown on Figure \ref{trees}.
%show a picture. 

Denote by $T_r$ a complete binary tree of height (root-leaf distance) $r$. The following structural lemma is the critical component 
of the proof of Theorem \ref{dmwtw}.

\begin{lemma} \label{dmwtwstruct}
Let $p$ be an arbitrary integer and let $H$ be an arbitrary connected graph of $2p$ vertices.
Then for any $r \geq \lceil log p \rceil$, $mw(T_r(H)) \geq (r+1-\lceil log p \rceil)p/2$
\end{lemma}
Before proving Lemma \ref{dmwtwstruct}, let us show how Theorem \ref{dmwtw}
follows from it.

{\bf Sketch proof of Theorem \ref{dmwtw}.}
First of all, let us identify the class ${\bf G}$. Recall that $P_x$ a path of $x$ vertices. Let $0 \leq y \leq 3$ be such that
$k-y+1$ is divided by $4$. The considered class ${\bf G}$ consists of all $G=T_r(P_{\frac{k-y+1}{2}})$ for
$r \geq 5\lceil log k \rceil$. It can be observed that the max-degree of the graphs of ${\bf G}$ is $5$
and their treewidth is at most $k$.

Taking into account that starting from a sufficiently large $r$ compared to $k$, 
$r=\Omega(log(n/k))$ can be seen as $r=\Omega(log n)$, the lower bound of Lemma \ref{dmwtwstruct}
can be stated as $mw(G)=\Omega(log n*k)$.
Finally, we observe that for bounded-degree graphs $mw(G)$ and $dmw(G)$ are linearly related and conclude
that a lower bound on $mw(G)$ implies the analogous lower bound on $dmw(G)$.
$\blacksquare$

The following lemma is an auxiliary statement for Lemma \ref{dmwtwstruct}.

\begin{lemma} \label{mincase}
Let $T$ be a tree consisting of at least $p$ vertices. 
Let $H$ be a connected graph of at least $2p$ vertices.
Let $V_1,V_2$ be a partition of $V(T(H))$ such that both
partition classes contain at least $p^2$ vertices. Then 
$T(H)$ has a matching of size $p$ with the ends of each 
edge belong to distinct partition classes. 
\end{lemma} 

%Now we are ready to prove Lemma \ref{dmwtwstruct}.    

{\bf Proof of Lemma \ref{dmwtwstruct}.}
The proof is by induction on $r$.
The first considered value of $r$ is $\lceil log p \rceil$.
After that $r$ will increment in $2$. In particular,
for all values of $r$ of the form $\lceil log p \rceil+2x$,
we will prove that $mw(T_r(H)) \geq (x+1)p$ and, moreover, for
each permutation $SV$ of $V(T_r(H))$, the required matching can be witnessed
by a partition of $SV$ into a suffix and a prefix of size at least $p^2$ each.
Let us verify that the lower bound $mw(T_r(H)) \geq (x+1)p$ implies the lemma.
Suppose that $r=\lceil log p \rceil+2x$ for some non-negative integer $x$.
Then $mw(G) \geq (x+1)p=((r-\lceil log p \rceil)/2+1)p>(r-\lceil log p \rceil+1)p/2$. Suppose $r=\lceil log p \rceil+2x+1$.
Then $mw(G)=mw(T_r(H)) \geq mw(T_{r-1}(H)) \geq (x+1)p=((r-\lceil log p \rceil-1)/2+1)p=
(r-\lceil log p \rceil+1)p/2$.

Assume that $r=\lceil log p \rceil$ and let us show the lower bound of $p$ on the matching width.
$T_r$ contains at least $2^{\lceil log p \rceil+1}-1 \geq 2^{log p+1}-1=2p-1 \geq p$
vertices. By construction, $H$ contains at least $2p$ vertices.
Consequently, for each ordering of vertices of $T_r$ we can
specify a prefix and a suffix of size at least $p^2$ (just choose a prefix
of size $p^2$). Let $V_1$ be the set of vertices that got to the prefix
and let $V_2$ be the set of vertices that got to the suffix. By 
Lemma \ref{mincase} there is a matching of size at least $p$ consisting of
edges between $V_1$ and $V_2$ confirming the lemma for the considered case.

Let us now prove the lemma for $r=\lceil log p \rceil+2x$
for $x \geq 1$. Specify the center of $T_r$ as the root  %why not a+1?
and let $T^1, \dots, T^4$ be the subtrees of $T_r$ rooted by the grandchildren of the
root. Clearly, all of $T^1, \dots, T^4$ are copies of $T_{r-2}$.
Let $SV$ be a sequence of vertices of $V(T_r(H))$.  Let $SV^1, \dots, SV^4$
be the respective sequences of $V(T^1(H)), \dots, V(T^4(H))$ `induced' by $SV$
(that is their order is as in $SV$).
By the induction assumption, for each of them we can specify a partition $SV^i_1,SV^i_2$ into a prefix and
a suffix of size at least $p^2$ each witnessing the
conditions of the lemma for $r-2$. Let $u_1, \dots, u_4$ be the last respective vertices
of $SV^1_1, \dots, SV^4_1$. Assume w.l.o.g. that these vertices occur in $SV$ in the
order they are listed. Let $SV',SV''$ be a partition of $SV$ into a prefix and 
a suffix such that the last vertex of $SV'$ is $u_2$. By the induction assumption we know
that the edges between $SV^2_1 \subseteq SV'$ and $SV^2_2 \subseteq SV''$ form a matching $M$ of size at least $xp$.
In the rest of the proof, we are going to show that the edges between $SV'$
and $SV''$ whose ends do not belong to any of $SV^2_1,SV^2_2$ can be used to form a matching $M'$
of size $p$. The edges of $M$ and $M'$ do not have joint ends, hence this will imply
existence of a matching of size $xp+p=(x+1)p$, as required. 

The sets $SV' \setminus SV^2_1$ and $SV'' \setminus SV^2_2$ partition
$V(T_r(H)) \setminus (SV^2_1 \cup SV^2_2)=V(T_r(H)) \setminus V(T^2(H))=
V([T_r \setminus T^2](H))$. Clearly, $T_r \setminus T_2$ is a tree.
Furthermore, it contains at least $p$ vertices. Indeed, $T^2$ (isomorphic to $T_{r-2}$) has $p$
vertices just because we are at the induction step and $T_r$ contains at least
$4$ times more vertices than $T^2$. 
So, in fact, $T_r \setminus T^2$ contains at least $3p$ vertices. 
Furthermore, since $u_1$ precedes $u_2$, the whole $SV^1_1$ is in $SV'$.
By definition, $SV^1_1$ is disjoint with $SV^2_1$ and hence it is a subset
of $SV' \setminus SV^2_1$. Furthermore, by definition,
$|SV^1_1| \geq p^2$ and hence $|SV' \setminus SV^2_1| \geq p^2$ as well.
Symmetrically, since $u_3 \in SV''$, we conclude that $SV^3_2 \subseteq SV'' \setminus SV^2_2$
and due to this $|SV'' \setminus SV^2_2| \geq p^2$.

Thus $SV' \setminus SV^2_1$ and $SV'' \setminus SV^2_2$ partition  
$V([T_r \setminus T^2](H))$ into classes of size at least $p^2$ each and the size
of $T_r \setminus T^2$ is at least $3p$. Thus, according to Lemma \ref{mincase},
there is a matching $M'$ of size at least $p$ created by edges between 
$SV' \setminus SV^2_1$ and $SV'' \setminus SV^2_2$, confirming the lemma,
as specified above $\blacksquare$

%\bibliographystyle{plain}
%\bibliography{KnowComp}

\appendix
\section{Transformation of a {\sc nrobp} into a uniform one}
Let $Z$ be the {\sc nrobp} being transformed and let $F$ be 
the function of $n$ variables realized by $Z$.  
Let $a_1,\dots, a_m$ be the non-leaf nodes of $Z$ being ordered
topologically. We show that there is a sequence $Z_{a_1}=Z, Z_{a_2}, \dots, Z_{a_m}$
such that each $Z_{a_i}$ for $i>1$ is a {\sc nrobp} of $F$ obtained from $Z_{a_{i-1}}$
by subdividing the in-coming edges of $a_i$ by adding at most $n$ nodes and $O(n)$ edges to each such
an in-coming edge. Moreover, the edges of any two paths $P_1$ and $P_2$ from the root of $Z_{a_i}$ to $a_i$ 
or to any node topologically preceding $a_i$ are labelled
with literals of the same set of variables. Observe that 
since each edge has only one head, say $a_j$, it is subdivided only once,
namely during the construction of $Z_{a_j}$. Hence the number of new added edges of $Z_{a_m}$ is $O(n)$ per edge of 
$Z$ and hence the size of $Z_{a_m}$ is $O(n)$ times larger than the size     
of $Z$.

Regarding $Z_{a_1}$ this existence statement is vacuously true so assume $i>1$
Denote by $AllVar(a_i)$ the set of all variables whose literals label 
edges of paths of $Z_{a_{i-1}}$ from the root to $a_i$. 

For each in-neighbour $a'$ of $a_i$, we transform the edge $(a',a_i)$ as follows.
Let $P$ be a path from the root of $Z_{a_{i-1}}$ to $a_i$ passing through $(a',a_i)$.
Let $x^1, \dots, x^q$ be the elements of $AllVar(a_i) \setminus Var(A(P))$.
We subdivide $(a',a_i)$ as follows. We introduce new nodes $a'_1, \dots, a'_q$ and
let $a_{q+1}=a$. Then instead $(a',a_i)$ we introduce an edge $(a',a'_1)$ carrying
the same label as $(a',a_i)$ (or no label in case $(a',a_i)$ carries no label).
Then, for each $1 \leq i \leq q$ we introduce two edges $(a'_i,a'_{i+1})$ carrying labels
$x^q$ and $\neg x^q$, respectively.

Let us show that the edges of any two paths $P_1$ and $P_2$ from the root of $Z_{a_i}$ to $a_i$ are labeled
with literals of the same set of variables. Let $a'$ be an in-neighbour of $a_i$ in 
$Z_{a_{i-1}}$. By the induction
assumption, any two paths from the root to $a'$ are labelled with literals of the same set of
variables. It follows that as a result any two paths from the root to $a_i$ passing through $a'$
are labelled by literals of the same set of variables, namely $AllVar(a_i)$. Since this is correct
for an arbitrary choice of $a'$, we conclude that in $Z_{a_i}$ any two
paths from the root to $a_i$ are labelled with $AllVar(a_i)$, that is with literals of the same set
of variables. Observe that the paths to the nodes of $Z$ preceding $a_i$ are not affected so
the `uniformity' of paths regarding them holds by the induction assumption. Regarding the new added nodes
on the subdivided edge $(a',a_i)$ the uniformity clearly follows from the uniformity of paths from the root
to $a'$.

To verify read-onceness of $Z_{a_i}$, let $P'$ be a path from the root to the leaf of $Z_{a_i}$.
Taking into account the induction assumption, the only reason why $P'$ may contain two
edges labelled by literals of the same variable is that $P'$ is obtained from a path $P$
of $Z_{a_{i-1}}$ by subdivision of an edge $(a',a_i)$ of this path. By construction the variables
of the new labels put on $(a',a_i)$ do not occur on the prefix of $P$ ending at $a_i$. Furthermore,
by definition of $AllVar(a_i)$ the variable $x$ each new label, in fact occurs in some path of $Z_{a_{i-1}}$
from the root to $a_i$ and hence, by the read-onceness, $x$ does not occur on any path starting from $a_i$.
It follows that the variables of the new labels do not occur on the suffix of $P'$ starting at $a_i$.
Taking into account that all the new labels of $(a',a_i)$ are literals of distinct variables, the read-onceness
of $P'$, and hence the read-onceness of $Z_{a_i}$, due to the arbitrary choice of $P'$, follow. Thus we know now that $Z_{a_i}$
is a {\sc nrobp}.

It remains to verify that $Z_{a_i}$ indeed realizes $F$. Let $P'$ be a path of $Z_{a_i}$ from the root to the
leaf. Then $A(P')$ is an extension of $A(P)$ of some path $P$ of $Z_{a_i}$. By the induction assumption,
any extension of $A(P)$ is a satisfying assignment of $F$, hence so is $A(P')$. 
Conversely, for each satisfying assignment $A$ of $F$ we can find a path $P$ of $Z_{a_{i-1}}$ such that 
$A(P) \subseteq A$. If an edge of path $P$ is subdivided then the new labels are opposite
literals on multiple edges. So, for every such multiple edge we can choose one edge carrying the literal
occurring in $A$ and obtain a path $P'$ such that $A(P') \subseteq A$. 

For the leaf node we do a similar transformation but this time add new labels on the in-coming edges of
the leaf so that the set of labels on each path from the root to the leaf is a set of literals of $Var(F)$.
A similar argumentation to the above shows that the resulting structure is indeed a uniform {\sc nrobp}
realizing $F$. Clearly the size of the resulting {\sc nrobp} remains $O(n)$ times larger than the size of $Z$. 
 
\section{Proofs of auxiliary statements for Lemma \ref{deepcover}}
{\bf Proof of Lemma \ref{onepos}.}
Assume the opposite that let $P$ be a path from the root to the leaf of $Y$
passing through $a$. It follows that in $A(P)$, $Var(a)$ occurs negatively.
Due to the monotonicity of $\phi(G)$, replacing $\neg Var(a)$ by $Var(a)$ in
$A(P)$ produces another satisfying assignment $A'$ of $\phi(G)$. Let $P^a$ be the prefix
of $P$ ending at $a$. Since $A(P^a) \subseteq A'$, by definition of a uniform {\sc nrobp},
there is a path $P'$ from $a$ to the leaf of $Y$ such that $A(P^a \cup P')=A'$.
Since $Var(a)$ occurs positively in $A'$ this is only possible if the successor of $a$
in $P'$ is its positive out-neighbour in contradiction to our assumption of its non-existence.
$\blacksquare$

To prove Lemma \ref{freeaprime}, we need an auxiliary statement. 
\begin{lemma} \label{fixfree}
Let $Y$ be a {\sc nfbdd} realizing $\phi(G)$ and let $a$ be a node of $Y$.
Let $P_1$ be a path from the root to $a$. Denote $Vert(A(P_1))$ by $Vrt$.
Let $A' \subseteq A(P_1)$ be the set of all negative literals of $A(P_1)$
and denote $Vert(A')$ by $Vng(P_1)$. Then $Free_a=V(G) \setminus (Vrt \cup N_G(Vng(P_1)))$.
%($N_G(Vng)$ denotes the open
%neighbourhood of $Vng$ in $G$).
\end{lemma}

{\bf Proof.}
Let $v \in Free_a$. Then $Y$ has a path $P_2$ from $a$ to the leaf such that
$Var(v)$ occurs negatively in $A(P_2)$. Due to read-onceness of $Y$,
$Var(v)$ does not occur in $A(P_1)$, hence $v \notin Vrt$. Assume that $v$ is a neighbour
of some $u \in Vng(P_1)$. By definition of $Y$, $A(P_1 \cup P_2)$ is a satisfying
assignment of $\phi(G)$ containing $\{\neg Var(u),\neg Var(v)\}$ which is a contradiction
since $\phi(G)$ contains a clause $(Var(u) \vee Var(v))$. Thus $v \notin N_G(Vng(P_1))$ and thus
we have verified that $Free_a \subseteq V(G) \setminus (Vrt \cup N_G(Vng))$.

Conversely, let $v \in V(G) \setminus (Vrt \cup N_G(Vng(P_1)))$. It follows that $Var(v)$ does not
occur in $A(P_1)$ and that $Var(v)$ does not occur in the same clause of $\phi(G)$ with any
of $Var(Vng(P_1))$. Consequently, there is a satisfying assignment $A'$ of $\phi(G)$ such that
$A(P_1) \subseteq A'$ and $Var(v)$ occurs negatively in $A'$: just assign positively
the rest of the variables. By definition of a uniform {\sc nrobp},
there is path $P_2$ from $a$ to the leaf of $Y$ such that $A(P_1 \cup P_2)=A'$.
Clearly $A(P_1 \cup P_2)=A(P_1) \cup A(P_2)$ and $Var(v)$ occurs negatively in $A(P_2)$.
Hence $v \in Free_a$ and thus we have confirmed that 
$V(G) \setminus (Vrt \cup N_G(Vng(P_1))) \subseteq Free_a$, completing the lemma. 
$\blacksquare$

{\bf Proof of Lemma \ref{freeaprime}.}
It is not hard to see that in each case the considered subset of $B$ is a subset of $Vert_{a'}$.
By Lemma \ref{fixfree}, it remains to set a path $P'$ from the root of $Y$ to $a'$ and
to verify that in each item the considered subset of $B$ does not have neighbours in 
$Vng(P')$ (as defined in Lemma \ref{fixfree}). Let $P$ be a path from the root to $a$ and
let $P'$ be a path obtained by appending $(a,a')$ to the end of $P$.
Clearly $Vng(P')$ is $Vng(P)$ plus, possibly, $Vert(Var(a))=v$ in case $a'$ is a negative out-neighbour
of $a$. Since $B \subseteq Free_a$, it follows from Lemma \ref{fixfree} that $B$ is not adjacent with
$Vng(P)$. Hence, it remains to verify that in each case the considered subset of $B$ is not adjacent with $v$.
This is certainly true in the first case because $B$ is an independent set and hence $B \setminus \{v\}$
is not adjacent with $v$. In the second case due to being $B$ a {\sc dis}, $v$ does not have neighbours in 
$B$ other than $w$ and hence $v$ is not adjacent with $B \setminus \{w\}$. In the third case either $a'$ is a positive 
out-neighbour of $a$ and hence $v \notin Vng(P')$ or $v$ is not adjacent to $B$ (otherwise we obtain the second
case). In any case, $B$ is not adjacent with $Vng(P')$.

\begin{comment}
In the first two items, $v$ is not a neighbour of the considered subset of $B$.
In case of $B \setminus \{v\}$, this is simply because $B$ is an independent set. In the second
case, $v$ does not have neighbours in $B$ other than $w$ due to being $B$ a {\sc dis}, hence the claim
is correct regarding $B \setminus \{w\}$. In the third case $v$ is adjacent to $B$ only if $Var(v)$ occurs positively in $A(P')$
(otherwise we have the second case), hence either $v \notin Vng(P')$ or $v$ is not a neighbour of $B$.
It follows that in all the considered cases it is enough to check that the absence of neighbours of $B$ in $Vng(P)$.
However, this follows from Lemma \ref{fixfree} becaause $B \subseteq Free_a$ and $P$ is a path from the root
of $Y$ to $a$. Hence the corollary follows. $\blacksquare$
\end{comment}

%simon 02088251277

{\bf Proof of Proposition \ref{pathincrease}.}
Indeed, $w((a,a')+{\bf P})=\sum_{P \in {\bf P}} w((a,a')+P)=
\sum_{P \in {\bf P}} (w(a,a')*w(P))=w(a,a')*\sum_{P \in {\bf P}} w(P)=
w(a,a')*w({\bf P})$, as required. $\blacksquare$

{\bf Proof of Lemma \ref{pathdecomp}.}
Suppose that $v \in B$. Let $P \in {\bf P}^B_a$. Clearly the element $a'$ following $a$ is an out-neighbour of $a$.
However, if $a'$ is the negative out-neighbour of $a$ then $Var(v)$ occurs negatively in $A(P)$ and hence $B$ does not
cover $P$, a contradiction. It remains to assume that $a'$ is the positive out-neighbour of $a$.
Hence, ${\bf P}^B_a$ can be represented as $(a,a')+{\bf P'}$ where ${\bf P'}$ is a set of paths starting at $a'$.
It remains to show that ${\bf P'} \subseteq {\bf P}^{B \setminus \{v\}}_{a'}$. Let $P' \in {\bf P'}$.
Then $A(P)=A((a,a')) \cup A(P')$ is covered by $B$ (here we admit a notational abuse identifying an edge with a path).
However, the only variable occurring positively in $A((a,a'))$ is $Var(v)$. It remains to assume that $Var(B \setminus \{v\})$
occur positively in $P'$, that is $P'$ is covered by $B \setminus \{v\}$. Thus we have proved the first statement.

Suppose $v \notin B$. Clearly, ${\bf P}^B_a$ is the union of all $(a,a')+{\bf P'}$ where $a'$ is an out-neighbour of $a$
and ${\bf P'}$ is some set of paths starting at $a'$. Let $P' \in {\bf P'}$. Then $A((a,a')+P')$ is covered by $B$, however $A((a,a'))$
is not covered by any subset of $B$. It remains to assume that $P'$ is covered by $B$ and hence ${\bf P'} \subseteq {\bf P}^B_{a'}$. $\blacksquare$

{\bf Proof of Lemma \ref{rwdecomp}.}
For the first item, notice that $Vert_{a'}=Vert_a \setminus \{v\}$
and that, due to being $B$ an independent set, no vertex of $B \setminus \{v\}$
is adjacent to $v$. It follows that that the neighbours of each $u \in B\setminus \{v\}$
in $Vert_{a'}$ are exactly the same as in $Vert_a$ and hence $ld_a(u)=ld_{a'}(u)$.
It follows that the factor contributed by each vertex of $B \setminus \{v\}$ to
$rw_{a'}(B \setminus \{v\})$ and to $rw_a(B \setminus \{v\})$ is the same. That is,
$rw_{a'}(B \setminus \{v\})=rw_a(B \setminus \{v\})=rw_a(B)/(1-2^{-(ld_a(v)+1)})$,
as required. 

For the second item, notice that, due to $B$ being a {\sc dis}, $v$ is not
a neighbour of any vertex of $B$ other than $w$. It follows that that the neighbours of each $u \in B\setminus \{w\}$
in $Vert_{a'}=Vert_a \setminus \{v\}$ are exactly the same as in $Vert_a$ and hence $ld_a(u)=ld_{a'}(u)$.
It follows that the factor contributed by each vertex of $B \setminus \{w\}$ to
$rw_{a'}(B \setminus \{w\})$ and to $rw_a(B \setminus \{w\})$ is the same. That is,
$rw_{a'}(B \setminus \{w\})=rw_a(B \setminus \{w\})=rw_a(B)/(1-2^{-(ld_a(w)+1)})$,
as required. On the other hand, $w$ has one neighbour less in $Vert_{a'}$ than in $Vert(a)$.
That is, $ld_{a'}(w)=ld_a(w)-1$. Clearly, $rw_{a'}(B)$ can be obtained by multiplying 
$rw_{a'}(B \setminus \{w\})$ by the factor contributed by $w$. That is 
$rw_{a'}(B)=rw_{a'}(B \setminus \{w\})*(1-2^{-ld_a(w)})=
 rw_a(B)*\frac{1-2^{-ld_a(w)}}{(1-2^{-(ld_a(w)+1)})}$. 

For the last item it is easy to see that the local degrees of vertices of $B$ are the same regarding $a'$
and $a$ and hence they contribute the same factor and the desired equality follows. 
$\blacksquare$

\section{Proofs of statements for Theorem \ref{dmwtw}}
The next lemma is an auxiliary statement needed for proving Lemma \ref{mincase}.
\begin{lemma} \label{matchontheway}
Suppose the vertices of $T(H)$ are partitioned into $2$ subsets $V_1$
and $V_2$. Let $L$ be a subset of vertices of $H$ such that $|L|=t$.
Suppose there are two copies $H_1$ and $H_2$ of $H$ such that for each
$u \in L$ the copies of vertex $u$ in $H_1$ and $H_2$
belong to different partition classes. 
Then $T(H)$ has matching of size $t$ with the ends of each edge lying
in different partition classes
\end{lemma}

{\bf Proof.}
Let $v_1$ and $v_2$ be the respective vertices of $T$ 
corresponding to $H_1$ and $H_2$. Let $p$ be the path between $v_1$
and $v_2$ in $T$. Then for each $u \in L$ there are two consecutive
vertices $v'_1$ and $v'_2$ of this path with respective copies $H'_1$
and $H'_2$ such that the copy $u'_1$ of $u$ in $H'_1$ 
belongs to the same partition class as the copy $u_1$ of $u$ in $H_1$ 
and the copy $u'_2$ of $u$ in $H'_2$ belongs 
to the same partition class as the copy $u_2$  of $u$ in $H_2$. 
By construction, $T(H)$ has an edge $\{u'_1,u'_2\}$ which we choose to correspond to $u$.
Let $L=\{u^1,\dots u^t\}$ and consider the set of edges as above corresponding to each $u^i$. 
By construction, both ends of the edge corresponding to each $u^i$ are copies
of $u^i$ and also these ends correspond to distinct partition classes. 
It follows that these edges do not have joint ends and indeed 
constitute a desired matching of size $t$ $\blacksquare$

{\bf Proof of Lemma \ref{mincase}.}
The proof is under assumption that $T$ contains \emph{exactly} $p$ vertices.
Indeed, otherwise, such a tree can be obtained by an iterative removal of 
the copies of $H$ associated with vertices having degree $1$. Clearly, any matching of the resulting restricted graph will
also be a matching of the original graph and the 
lower bound on the sizes of the partition classes
will be preserved as well.

Assume first that each copy of $H$ corresponding to a vertex of $T$
contains vertices of both partition classes. Since $H$ is a connected
graph, for each copy we can specify an edge with one end in $V_1$ and
the other end in $V_2$. These edges belong to disjoint copies of $H$,
hence none of these edges have a common end. Since there are $p$ copies
of $H$, we have the desired matching of size $p$.

Assume now that there is a vertex $u$ of $T$ such that the copy $H_1$ of $H$
corresponding to $u$ contains vertices of only one partition class. 
Assume w.l.o.g. that this class is $V_1$. Then there is a vertex
$v$ of $T$ such that the copy $H_2$ of $H$ corresponding to $v$ contains at least
$p$ vertices of $V_2$. Indeed, otherwise, at most $p-1$ vertices per $p$
copies will not make $p^2$ vertices altogether. Let $L$ be the set of vertices of $H$ 
whose copies in $H_2$ belong to $V_2$. By assumption, all the copies of $L$ 
in $H_1$ belong to $V_1$. By Lemma \ref{matchontheway},
$H_1$ and $H_2$ witness the existence of matching of size $p$ with ends of each
edge belonging to distinct partition classes. $\blacksquare$ 

In order to prove Theorem \ref{dmwtw}, we need an auxiliary proposition.

\begin{proposition} \label{dmw1}
For a graph $G$ with maxdegree $c$, $dmw(G) \geq mw(G)/(2c^2+2c+1)$. 
%Recall that $dmw(G)$ denotes the {\sc dmw} of $G$ and $mw(G)$ denotes the matching width of $G$.
\end{proposition}
{\bf Proof.}
For each ordering of vertices of $G$ take the partition witnessing $mw(G)$ and let $M$
be a witnessing matching of size $mw(G)$. Let $\{u_i,v_i\}$ be an edge of $M$.
It is not hard to see that the number of vertices $v$ whose
open neighbourhood intersects with that of $\{u_i,v_i\}$ is at most $(2c+2)c$. 
Indeed, $|N[u_i,v_i]| \leq 2c+2$. If for some vertex $v$, 
$N[v] \cap N[u_j,v_j] \neq \emptyset$ then $v \in N[N[u_j,v_j]]$.
Clearly, $|N[N[u_j,v_j]]| \leq (2c+2)c$ as required.

Now, let us create a distant matching $M^*$ out of $M$.
Take $\{u_1,v_1\}$ to $M^*$ and remove it from $M$ together with at most $2c^2+2c$ pairs
whose open neighborhood may intersect with $N[u_1,v_1]$. Until $M$ is not empty
take the survived $\{u_i,v_i\}$ of the smallest index $i$ and perform the same operation.
It clearly follows by construction that $M^*$ is a distant matching. Let us 
compute its size. On each step the number of pairs removed from $M$ is at most $2c^2+2c+1$,
so the number of iterations of adding pairs to $M^*$ and hence the number of such pairs is
at least $mw(G)/(2c^2+2c+1)$. 

We conclude that for each permutation of vertices of $G$ there is a partition witnessing
the desired distant matching, as required. $\blacksquare$ 

{\bf Proof of Theorem \ref{dmwtw}.}
First of all, let us identify the class ${\bf G}$. Recall that $P_x$ a path of $x$ vertices. 
Further on, let $0 \leq y \leq 3$ be such that
$k-y+1$ is divided by $4$.
The considered class ${\bf G}$ consists of all $G=T_r(P_{\frac{k-y+1}{2}})$ for $r \geq 5\lceil log k \rceil$.

Let us show that the treewidth of the graphs of ${\bf G}$ is bounded by $k$. 
Consider the following tree decomposition of $G=T_r(H=P_{\frac{k-y+1}{2}})$. The tree is $T_r$.
Consider $T_r$ as the rooted tree with the centre being the root. The bag of each vertex includes
the vertices of the copy of $H$ associated with this vertex plus the copy of the parent (for a non-root vertex). 
The properties of tree decomposition can be verified by a direct inspection.
The size of each bag is at most $k-y+1$, hence the treewdith is at most $k-y \leq k$.

Observe that max-degree of the graphs of ${\bf G}$ is $5$. Indeed, consider a vertex
$v$ of $G \in {\bf G}$ that belongs to a copy of $H$ associated with a vertex $x$ of some $T_r$.
Inside its copy of $H$, $v$ is adjacent to at most $2$ vertices. Outside its copy
of $H$, $v$ is adjacent to vertices in the copies of $H$ associated with the neighbours
of $x$, precisely one neighbour per copy. Vertex $x$ is adjacent to at most $3$ vertices
of $T_r$. It follows that $v$ has at most $3$ neighbours outside its copy of $H$.

It follows from Proposition \ref{dmw1} that {\sc dmw} and the matching width of graphs of ${\bf G}$
are linearly related. Therefore, it is sufficient to obtain the desired lower bound on 
the matching width. This is done in the next paragraph.

Let us reformulate the lower bound of $mw(G)$ in terms of $log n$ and $k$ where $n=V(G)$.
Notice that $p$ used in Lemma \ref{dmwtwstruct} can be expressed as $(k-y+1)/4$.
Hence, the lower bound on the matching width can be seen as
$(r-\lceil log(\frac{k-y+1}{4}+1) \rceil)*(k-y+1)/8$. This lower bound can be immediately simplified
by noticing that by the choice of $k$ and $y$, $(k-y+1)/8 \geq k/16$ 
and $\lceil log(\frac{k-y+1}{4}) \rceil \leq \lceil log k \rceil$. Hence,
$(r-\lceil log k \rceil+1)k/16$ can serve as a lower bound on $mw(G)$. 
To draw the connection between $n$ and $r$, notice that $n=(2^{r+1}-1)(k-y+1)/2$.
It follows that $r+1=log(\frac{n}{(k-y+1)/2}+1)$. In particular, it follows that
$r+1 \geq log n-log k \geq log n -\lceil log k \rceil$. 
It follows that $r+1$ in the lower bound can be replaced by $log n -\lceil log k \rceil$ and the new
lower bound is $(log n-2 \lceil log k \rceil)k/16$. Consequently, for $log n \geq 5 \lceil log k \rceil$
the lower bound can be represented as $(log n*k)/32$ which is the form needed for the theorem.
It remains to observe that $r \geq 5\lceil log k \rceil$ implies $log n \geq 5\lceil log k \rceil$.
By the above reasoning, $r \geq 5\lceil log k \rceil$ implies $log(\frac{n}{(k-y+1)/2}+1) \geq 5\lceil log k \rceil$.
By our choice of $k \geq 50$, $log(n/20+1) \geq log(\frac{n}{(k-y+1)/2}+1)\geq 5\lceil log k \rceil$. By construction of $G$ and the choice of $r$,
$n \geq 2^{r+1}-1 \geq k^5-1 \geq k$, the last inequality follows from the choice of $k$, hence $n \geq 50$.
In particular, it follow that $n \geq n/20+1$. Hence $log n \geq log(n/20+1)\geq 5\lceil log k \rceil$.
$\blacksquare$
\end{document}